# Fairness-Oriented Semi-Chaotic Genetic Algorithm-Based Channel Assignment Technique for Nodes Starvation Problem in Wireless Mesh Network

Fuad A. Ghaleb[1][2], Bander Ali Saleh Al-rimy[3], Maznah Kamat[1*], Mohd. Foad Rohani[1], Shukor Abd Razak[1]

[1] School of Computing, Faculty of Engineering, Universiti Teknologi Malaysia, 81310, Johor, Malaysia
[2] Department of Computer and Electronic Engineering, Sana'a Community College, Sana'a 5695, Yemen
[3] Faculty of Business and Technology, Unitar International University, Petaling Jaya, Selangor, Malaysia

Corresponding author: Maznah Kamat(e-mail: kmaznah@utm.my), Fuad A. Ghaleb (e-mail: fuadeng@ gmail.com)

**ABSTRACT**
Multi-Radio Multi-Channel Wireless Mesh Networks (WMNs) have emerged as a scalable, reliable, and agile wireless network that supports many types of innovative technologies such as the Internet of Things (IoT) and vehicular networks. Due to the limited number of orthogonal channels, interference between channels adversely affects the fair distribution of bandwidth among mesh clients, causing node starvation in terms of insufficient bandwidth, which impedes the adoption of WMN as an efficient access technology. Therefore, a fair channel assignment is crucial for the mesh clients to utilize the available resources. However, the node starvation problem due to unfair channel distribution has been vastly overlooked during channel assignment by the extant research. Instead, existing channel assignment algorithms either reduce the total network interference or maximize the total network throughput, which neither guarantees a fair distribution of the channels nor eliminates node starvation. To this end, the Fairness-Oriented Semi-Chaotic Genetic Algorithm-Based Channel Assignment Technique (FA-SCGA-CAA) was proposed in this paper for Nodes Starvation Problem in Wireless Mesh Networks. FA-SCGA-CAA optimizes fairness based on multiple-criterion using a modified version of the Genetic Algorithm (GA). The modification includes proposing a semi-chaotic technique for creating the primary chromosome with powerful genes. Such a chromosome was used to create a strong population that directs the search towards the global minima in an effective and efficient way. The outcome is a nonlinear fairness oriented fitness function that aims at maximizing the link fairness while minimizing the link interference. Comparison with related work shows that the proposed FA_SCGA_CAA reduced the potential nodes starvation by 22% and improved network capacity utilization by 23%. It can be concluded that the proposed FA_SCGA_CAA is reliable to maintain high node-level fairness, while maximizing the utilization of the network resources, which is the ultimate goal of many IoT and vehicular network applications.

## I. INTRODUCTION

Wireless mesh networks (WMNs) enable flexible and robust connectivity for various applications, such as broadband internet service for home networks, healthcare, smart grids, Internet of Things (IoT) and intelligent transportation systems [1] [2]. The ability to use different radio technologies, including IEEE 802.11 (a/b/g/n) and 802.16, makes WMN flexible enough to support many manufacturing standards [3]. Moreover, the ability to work with other technologies such as ad-hoc wireless networks makes it easy to integrate WMNs with existing infrastructure and provide nodes with the ability to share and exchange data over the Internet. Client meshing is one of the important characteristics that distinguish WMNs from conventional wireless ad-hoc networks [4]. Such characteristic enables peer-to-peer communication among mesh nodes. Therefore, the nodes on the backbone provide the routing and configuration functionalities to the end-users [4]. Furthermore, WMN nodes are characterized as dual-functioning, such that they play both client and router roles by automatically establishing and maintaining connectivity among themselves [3]. Such property contributes to achieving reliable, low-maintenance, low-cost and robust mesh networks.

As opposed to Point-to-Point (PTP) communication that is used by traditional ad-hoc networks, WMNs use multipoint to multipoint (MTM) communication to increase network scalability, reliability, and capacity by enabling a mesh node to communicate with more than one other mesh node simultaneously[5]. WMNs consist of three main components, namely mesh routers, gateways, and clients [1]. Mesh routers work as a backbone that connects mesh clients and gateways. Gateways are mesh nodes that interconnect the WMN with other networks and services such as the Internet, data-centers, and servers. Mesh clients are nodes that end-users use to connect to the WMN. These nodes may be laptops, mobiles, vehicles, health care appliances, and any other IoT devices [6]. A mesh client reaches the resources or services by connecting to WMN via mesh routers, which in turn, redirects the traffic from/to the gateway. Mesh routers utilize multiple radio interfaces with multiple channels. Unlike traditional wireless routers that have a single radio interface, mesh routers use multiple radios to decrease the interference between co-located communication links as well as improve the throughput, connectivity, and capacity of the network. To decrease such interference, the co-located links need to use none-overlapping (orthogonal) channels. However, the limited number of orthogonal channels makes the interference between adjacent links inevitable. Therefore, effective channel assignment is key to ensure high network throughput, connectivity, and capacity [3, 7].

Channel assignment algorithms play an important role in improving the connectivity, throughput, and capacity of WMNs. These algorithms aim at finding an optimal distribution of the channels among the co-located links to maximize the utilization of network resources. Channel assignment algorithms try to allocate the required number of channels to each region in such a way that interference is reduced, and the frequency spectrum is used efficiently [3]. This problem is usually formulated as a graph coloring problem, which is naturally an NP-hard (a nondeterministic polynomial-time) problem whose optimal solution might not exist [8]. This is because in most practical situations, there may be insufficient orthogonal channels to ensure interference-free channel assignments. A large number of mesh devices may share a single common channel to reduce network interference. However, increasing the number of adjacent devices that use the same channel adversely affects the connectivity, bandwidth, and capacity of WMNs, which in turn causes nodes starvation due to unfair sharing of channels among mesh links. Such node starvation decreases the fair sharing of network resources, and consequently, its performance.

The node starvation problem occurs when the surrounding links unable to support the required bandwidth of the adjacent clients due to links interference. Fig. 1 illustrates an example where the node starvation problem occurs. As shown in Fig.1 (a), the capacity in terms of total supported bandwidth of mesh router A is 16Mbps while the required bandwidth is at 6Mbps. In the ideal situation, the available bandwidth can satisfy the requirements of mesh clients. However, Fig 1(b) shows that mesh router A unable to support more than 3Mbps bandwidth after channel assignment due to interference among adjacent links.. This results in the nodes starvation problem. Thus, a fair channel assignment algorithm should guarantee the equitable distribution of the bandwidth among the links such that all clients are served fairly. That is, the data rate of a link after the channel assignment should fulfill the data rate requirements of the mesh nodes connected to it. Therefore, a fair channel assignment should aim at ensuring that all links in the network can achieve a data rate that is suitable for all relevant nodes. Hence, the data rate of each link after the channel assignment should be consistent with the designated data rate of the link.

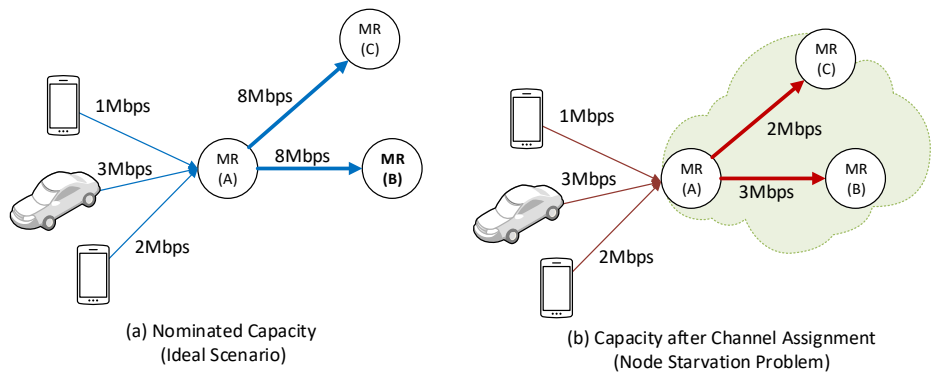

FIGURE 1. Node Starvation Problem

Over the last decade, many solutions have been suggested to address various channel assignment problems [1, 3-5, 7, 9-16]. Because channel assignment problems are NP-Hard, it is not feasible to use deterministic algorithms to find the optimal solutions in a finite amount of time. This is why most of the existing solutions used heuristics algorithms to approach the channel assignment problem. Unfortunately, heuristic algorithms may lead to inefficiencies, whereby the appropriate solutions may not be found within a reasonable period of time due to being trapped in local minima causing the unfair distribution of the channels in the network. Metaheuristic algorithms such as Genetic algorithms and Tabu search address this shortcoming by by approaching the global minima using the concept of natural selection and the evolutionary theory to [17]. That is, a new solution is resulting from combining two good solutions. However, the existing Genetic Algorithm (GA) based channel assignment algorithms [18-23] have overlooked the fairness issue. Most of those algorithms try to minimize the total sum of the links interferences, which unfortunately, does not guarantee fair channel distribution and cannot prevent node starvation problems. Although fair channel assignment was the subject of several studies in the recent few years [2, 9, 24-31], most of those solutions are not feasible due to the use of the heuristic approaches. In addition, many of those approaches focus on flow fairness, whereby the channel assignment algorithm distributes the available non-overlapping channels in such a way that flows (paths) in the networks have equal data rates. On the other hand, node fairness, which can significantly improve the fairness, has received low research attention [24]. Node fairness can be achieved by equitable distribution of bandwidth over the nodes. Consequently, equitable distribution of the channels should make each node end up with the desired bandwidth. In Fig. 1(b), for example, node A requires minimum bandwidth of 6Mbps but it is given only a total of 3Mbps. Ignoring node fairness leads to node starvation, which adversely affects the capacity and throughput of WMNs.

To this end, this paper proposes a channel assignment technique called the Fairness-Oriented Semi-Chaotic Genetic Algorithm-Based Channel Assignment Technique (FA-SCGA-CAA) that addresses the issue of node starvation through fair (equitable) channel distribution among all mesh nodes in the WMN. The channel assignment problem has been formulated as an optimization problem with two objectives, minimizing the interference and maximizing node fairness. Because interference and fairness are not linearly correlated, this study introduces a new non-linear fitness function that aims at minimizing the interference while maximizing the supported bandwidth to ensure fair distribution of the non-overlapping channels and guarantee the required bandwidth for each mesh client. Unlike existing channel assignment solutions that minimize the global interference to improve the fairness, this study aims at minimizing the link interference such that the node fairness is maximized. This study takes into account the node fairness when distributing the channels among the co-located links in WMNs. FA-SCGA-CAA optimizes fairness based on multiple-criterion using a modified version of the Genetic Algorithm (GA). The modification includes proposing a semi-chaotic technique for creating the primary chromosome with powerful genes. Such a chromosome was used to create a strong population that directs the search towards the global minima in an effective and efficient way. The outcome is a nonlinear fairness oriented fitness function that aims at maximizing the link fairness while minimizing the link interference. The contribution of this paper is five-fold.

1) The Fairness-Oriented Semi-Chaotic Genetic Algorithm-Based Channel Assignment Algorithm is proposed to maximize link fairness while minimizing link interference.
2) A semi-chaotic genetic-based technique is proposed to create a diverse population with informative features that converges at the best solution and avoids being trapped in the local minima. The semi-chaotic technique is proposed to address two main issues of Genetic Algorithms to speed up the convergence process of the algorithm and to increase the diversity of the searched solutions so as to find the best feasible solution.
3) The problem of fair channel assignment is formulated as an optimization problem which entails a fitness function that combines several factors representing the network topology, link capacity, and required bandwidth/throughput to minimize link interference while maximizing link fairness.
4) A new nonlinear fitness function is proposed to integrate both interference and fairness in one fitness function for minimizing link interference while maximizing link fairness that is directly reflected to improving client fairness.
5) Extensive experimental evaluations were conducted to measure the performance of the proposed technique and compare it with existing solutions.

The rest of this paper is organized as follows: The related works are reviewed in Section II. The proposed model is elaborated in Section III. The performance evaluation and the experimental setup are presented in Section IV. The results are illustrated and discussed in Section V. The paper is concluded in Section VI.

## II. RELATED WORK

The channel assignment problem in a multi-radio wireless mesh network has been the subject of many recent studies [3, 5, 9-13, 18, 24]. Many methods were used in those solutions such as graph-based [32, 33], optimization-based [18, 27, 34], and artificial intelligence-based [18-20, 35] techniques. A detailed review of those methods can be found in the following surveys [3, 7, 36]. Most of those studies aimed at minimizing global network interference. The links interferences were estimated using either protocol [37] or physical interference [38] models. The hypothesis behind those solutions was to reduce global network interference leads to improving the utilization of network capacity, throughput, goodput, delay, among other desired characteristics [3, 7, 36]. Although many of these solutions minimized global interference and accordingly improved the performance of the network, such solutions suffer from both links and nodes starvation problems due to the unfair distribution of the channels among the links. Nodes starvation problem occurs when a node tries to use a link with high interference. To solve this issue a fair distribution of the non-overlapped channels is required. Fair channel distribution has been the focus of several recent studies [2, 9, 24-31].

Fairness is defined by many researchers as the equal distribution of the resources among equal nodes [2, 26]. This broad definition of fairness has led many open challenges in channel assignment algorithms that need to be addressed such as fair bandwidth, fair throughput, fair goodput, fair access to the channel, and load balancing. The interference among communication channels is a major challenge that impedes the fair utilization of WMN resources and causes the node starvation problems.

Fairness in wireless networks can be categorized based on several criteria such as granularity, time, resource type, and access mode. From the granularity perspective, fairness can be categorized into system-wide fairness [24, 39, 40], per-flow fairness [26, 29], per-link-fairness [9, 28] and per-node- fairness [24, 26]. Such categorization determines the level at which fairness is achieved. System fairness is viewed from the perspective of the whole system, which is achieved when all mesh nodes attain individual fairness [24]. Per-flow fairness refers to equal bandwidth for all traffic flows arriving at the gateway [29], which is achieved by assigning non-overlapping channels to the links before allocating those links to flows based on the interference model [26]. Per-link fairness based solutions aim at ensuring equal distribution of the bandwidth among links. The per-link fair channel allocation tries to prevent nodes starvation phenomena that could happen due to the interference. Therefore the solutions that try to achieve per-link fairness aimed at minimizing the interference among links. Per-node fairness based solutions aim at ensuring that all nodes in the network have obtained equal access opportunity to the network. This can be achieved if the channel assignment algorithm takes the traffic demands of each individual link as a requirement during channel assignment to achieve fairness and prevent nodes starvation problems. This paper focuses on per-link fairness in order to achieve the per-node fairness. To the best of our knowledge, such granularity of fairness has not received enough investigation yet.

In their study, Qu, et al. [24], proposed a channel assignment algorithm to prevent the flow starvation problem. An interference model was embedded into the channel assignment algorithm to better estimate interference on the links so as to eliminate border effect and flow starvation. However, the algorithm was designed for single radio mesh networks. Moreover, such a solution does not consider the node starvation as it aims only to achieve per-flow fairness.

Bakhshi and Khorsandi [28] used Integer Linear Programming (ILP) to develop a dynamic channel assignment algorithm in order to achieve flow fairness. However, the solution defines fairness as a function of the number of accepted demands with a source-destination pairs over the specific threshold. Hence, fairness is achieved only when the number of transmission demands approaches a particular number. This approach can improve network capacity if the number of transmission demands is known prior to channel assignment. However, thus approach is scenario-specific, dynamic and cannot be defined in advance. Beheshtifard and Meybodi [41] devised an adaptive scheme based on learning automata to maintain channel assignment when network traffic demands dynamicaly changes. However, the scheme lacks a mechanism that ensures fair allocation of the channels among links.

Ghaleb, et al. [9] proposed a channel assignment algorithm based on weighted link ranking to achieve an equitable distribution of the orthogonal channels. The equitability (fairness) was achieved by employing multiple criteria like proximity from the gateway, expected traffic, and link capacity, to rank the mesh links. Then, the non-overlapping channels were assigned accordingly. However, such an approach lacks proper fairness measures that can effectively evaluate the level of fairness achieved.

In their study, Liu, et al. [42] proposed a genetic algorithm based routing algorithm combined with a channel assignment technique that aims at maximizing the minimum flow rate so as to improve flow fairness. However, maximizing the number of flows does not guarantee that node starvation will be prevented due to the inconsideration of per link interference during channel assignment. Genetic algorithm was used to search for the chromosome with the highest number of flows. In their solutions, the chromosomes were represented and evaluated based on the power level of the channels. However, power level has spatiotemporal characteristics that leads to unstable and short-term channel assignment and thus, such solution neither prevents link starvation nor flow starvation.

To sum up, many recent studies investigated fair channel assignment solutions. However, there are two main drawbacks of the extant research, which can be described as follows. First, most of those studies simply minimizing the total interference in the network to achieve effective utilization of network resources. Such solutions lead to fair starvation of nodes and links which is not the goal of fairness in channel assignment. The common hypothesis among these solutions states that reducing the total interference can improve network performance, which directly achieve fairness among mesh nodes. This hypothesis is inaccurate because reducing the total interference does not necessarily ensure equal distribution of channels in the network. In addition, fairness was represented in terms of total power level, total interference or total throughput of the network. Moreover, the extant studies aimed at equal distribution of network resources (such as the bandwidth) among nodes which does not necessarily lead to fairness and thus does not prevent node starvation problem. The equitable distribution of the resources among all nodes is the ultimate goal of a fair channel assignment algorithm to resolve the node starvation problem. That is, equal nodes only should receive equal resources. The resources should be distributed based on the requirements, which are not necessarily equal. The second drawback of the existing solutions is the used of heuristic methods to assign the channels. Unfortunately, heuristic techniques are not appropriate solutions as they are scenario dependent that cannot be generalized. Metaheuristics techniques in the other hand such as Genetic algorithms seem to be promising to approach channel assignment algorithm in WMN due to the smaller search scope as it searches in a controlled population sample. However, randomly generating the population samples may lead to a premature solution since the search can be directed to local minima.

This study addresses this issue by proposing a fair channel assignment algorithm with more granularity. Links and nodes fairness have been represented based on the impact of interference on their expected bandwidth. Thus, a fairness-oriented fitness function was proposed and integrated into the genetic algorithm based channel assignment, which aims at maximizing node and link fairness, while minimizing interference.

## III. The Proposed Semi-Chaotic Genetic Algorithm Based Channel Assignment

In this study, the problem of channel assignment is formulated as an optimization task that maximizes link fairness so as to avoid node starvation problems. Thus, we have proposed a semi-chaotic genetic algorithm to solve such a problem. The proposed algorithm aims at finding the most effective solution that achieves node fairness and addresses the link starvation problem. The Genetic Algorithm (GA) is a metaheuristic algorithm inspired by the evolution theory and natural selection. GAs are adaptive search techniques that can find the optimal global solution by manipulating and recursively generating a new population of solutions from an initial population space. GA is used for combinatorial optimization problems, namely the NP-optimization problems (NPO), since they search from one population of points in search space to another and tend to focus increasingly on areas with deeper minima [43]. The proposed algorithm comprises five steps as follows: network representation, semi-chaotic based initial population creation, fairness-aware individual evaluation, parent's selection, children generation or offspring. The channels are represented by genes. Hence, the number of genes is equal to the number of non-overlapping channels. Therefore, the chromosome represents a solution, which is a series of channels (genes) assigned to the radios in the network. Figure 1 shows the flowchart of the proposed algorithm.

As shown in Fig. 1, there are seven main steps in the algorithm. The first step is network representation. The WMN is represented as a graph so as to represent the interference between the links as elaborated in Section A. Then, the initial population has been created by developing a new semi-chaotic method, in which the genes of the primary chromosome is created using multi-criterion link ranking-based channel assignment algorithm [9]. In doing so, the initial population inherits powerful features from the father (the primary chromosome). The multi-criterion links ranking based channel assignment algorithm is described in Section B. To create diversity in the population and prevent the convergence to a local minimum, a chaotic based algorithm is used to create the initial population. Next, the individuals in the population are evaluated using a novel nonlinear fitness function. The optimization objectives are fuzzed into one representative nonlinear function so-called fairness-aware interference minimization function. This function correlates the interference to fairness by formulating the link fairness as a function of interference and the data rate. Thus, individuals or the chromosomes are evaluated according to fairness conditions. That is, if an individual is found to have an acceptable fairness index, the objective is achieved, and the algorithm is stopped. Otherwise, the algorithm continues to the next step. A detailed description of this function is found in Section C. The next step is the selection of the parents for the new generation. The best individuals with high values of fairness are selected for the offspring. Then, the semi-chaotic process is followed to create the new generation from the selected parents' chromosomes. For creating the children of the new generation, a semi-chaotic mutation approach is developed in which the genes (channels) with high fairness values are chosen for crossover, while the genes with low fairness is replaced using the semi-chaotic process. Finally, to maintain the diversity among the new generation, the genes with high fairness values are fixed while the genes with low fairness values are mutated. The steps from four to seven are iterated until the convergence in the global minima is obtained, or the maximum number of iteration is reached. Section D elaborates the last three steps in more detail.

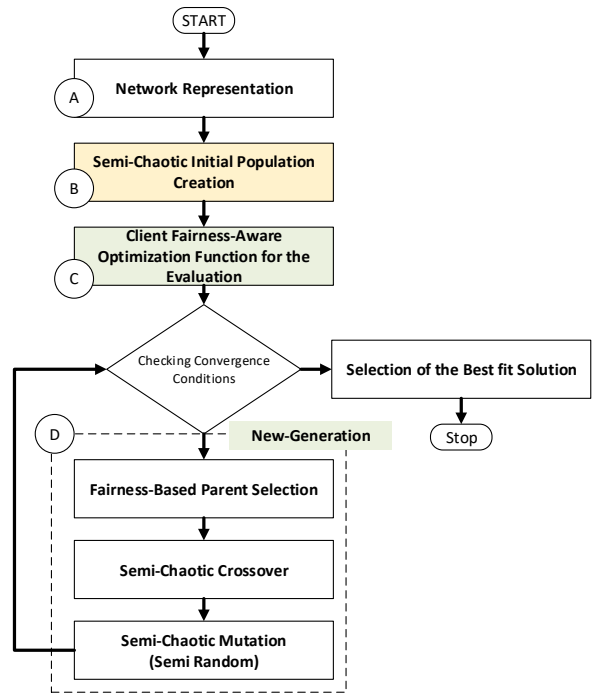

**FIGURE 1. Flow chart of the suggested approach**

## A. Network Representation

The wireless mesh network (WMN) is represented by a graph that can represent the interconnection of all nodes connected in the network. The graph consists of two main components, namely vertices, and edges. A vertex can be a mesh router, mesh client, or mesh gateway. Meanwhile, the edges represent the links between the vertices. A wireless link can be defined as a dedicated connection between two radio interfaces on two different nodes. The radio interfaces that form the links share the same radio characteristics, such as channel frequency, bandwidth, speed, and encoding. A link length is the Euclidian distance between the positions of the radio interfaces that forms that link. Two links are considered conflicted (and can also be partially overlapped) if they use the same channel and the distance between them is less than the channel interference range. The data rate of the links is affected by two factors the amount of the interference and the link length. A good channel assignment algorithm can result in low interference (internal interference) and high fairness. That is, the interference should be reduced and distributed fairly on the network so that nodes don't starve to obtain equitable bandwidth.

## B. Semi-chaotic Initial Population Formation

Forming the initial population is an important step in genetic algorithms that directly affects the quality of the results. There are two approaches for the initialization, namely heuristic and random. The heuristic approach may direct the GA to fast converge to a local optimum, while the random based approach may slow down the convergence. Therefore, in this study, a semi-chaotic based population formation technique is proposed. In this technique, firstly, the primary chromosome which constructs the entire population has been created using a multi-criterion channel assignment algorithm. The primary chromosome was created using strong genes so that the entire population inherits those powerful genes. Thus, the conference will be efficient, and Fig. 2 shows the flow chart of the primary chromosome with effective solution. The multi-criterion channel assignment algorithm was used to create the primary chromosome of the initial population.

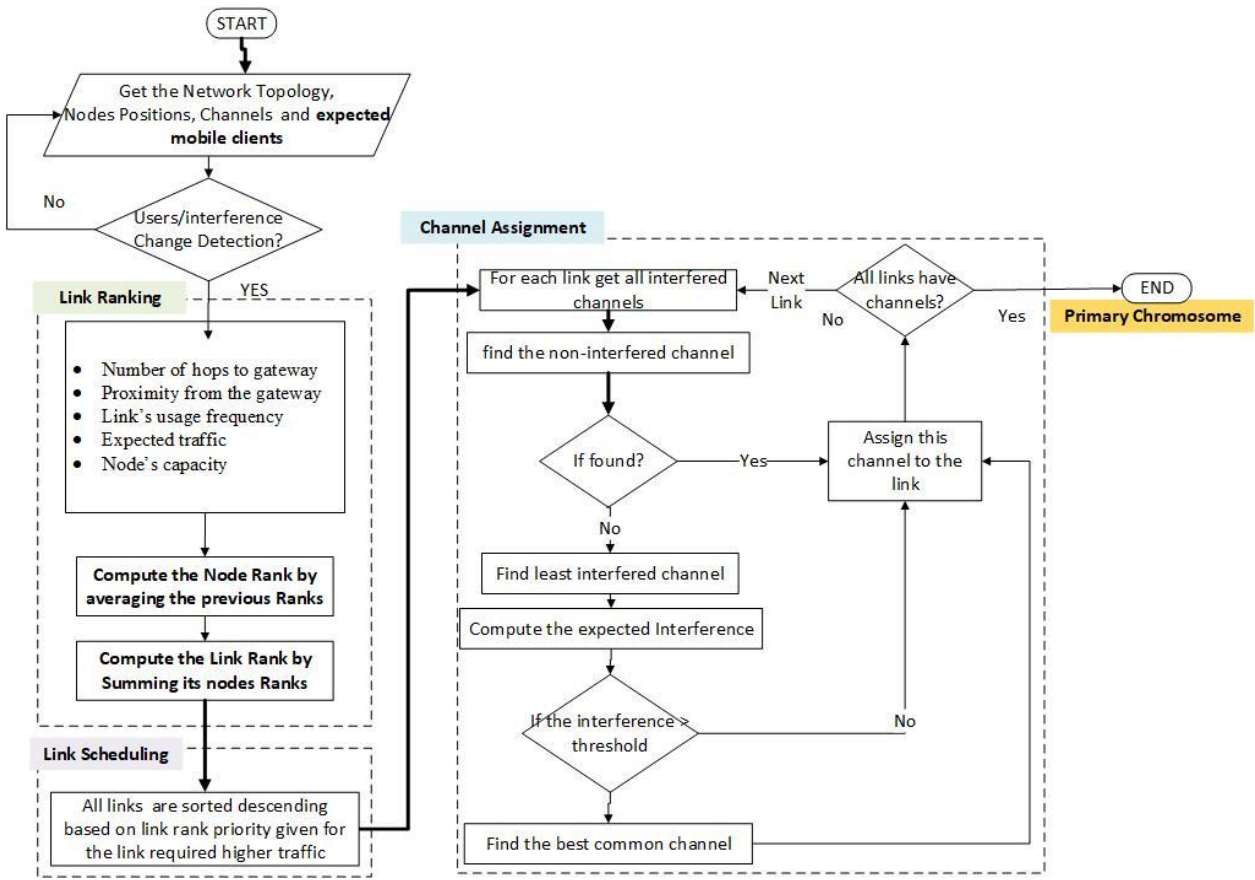

**FIGURE 2.** Flow chart for the generation of the primary chromosome

The multi-criterion channel assignment algorithm is a heuristic algorithm that uses multiple criteria derived from network topology and the expected traffic patterns used in the proposed channel assignment algorithm. As shown in Fig. 2, the algorithm is composed of three main steps, namely link ranking, link scheduling, and channel assignment. In the link ranking step, five criteria are used to rank the nodes as follows: a number of hops to the gateway, proximity from the gateway, usage frequency, and capacity. The node rank is normalized, and score for each node is given. The link rank is obtained by summating up the scores of the nodes that form the links. The next step is link scheduling; in which the links are sorted in descending order. Thus, the channel assignment algorithm starts distributing the non-overlapping channels over the links that have high ranks. Finally, the channel assignment is performed to obtain the primary chromosome to create the initial population. The process of obtaining the primary chromosome is as follows (See Fig. 2). Firstly, for each link in the graph, the list of all interfered channels is obtained. Then, the list of all non-overlapping channels that can be used for the link is obtained. If there is any such channel, it is assigned to the link. The algorithm then moves to the next important link. If the non-overlapping channels are not available for the current link, the least interfering channel is assigned.

The least interfering channel is calculated in every iteration of channel assignment. To estimate the expected interference, the mesh network is represented as a conflicting graph where the vertices are the links, and the edge is the shortest distance between any two nodes from adjacent links. Fig. 3 shows the difference between the link graph (Fig. 3 (a)), the interference graph (Fig. 3 (b)), and the conflicting graph (Fig. 3 (c)). The link graph represents the mesh network, which is used for computing the nodes and links ranking. The interference graph is used to obtain the interfered radios which used to form the conflicting graph. The conflicting graph is used to obtain the list of the interfered links during channel assignment. It is also used to calculate the actually expected bandwidth that the link can support after channel assignment. Thus, the conflicting Graph is used to evaluate the effectiveness of the generated solutions.

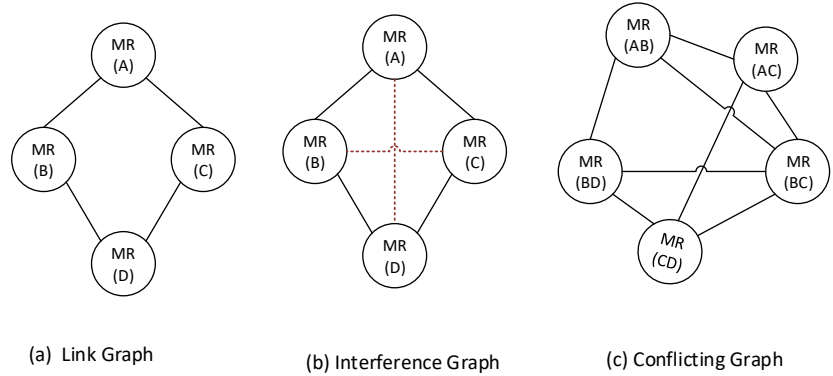

(a) Link Graph   (b) Interference Graph   (c) Conflicting Graph

**FIGURE 3: Link Graph, Interference Graph, and Conflicting Graph**

It worth noting that the channel assignment algorithm in this stage uses a threshold for the acceptable interference for assigning a channel to a link. If the resulting interference is not accepted, the algorithm assigns a common channel as no other option in this case. That is, the resulting bandwidth of the link after channel assignment should not be less than the bandwidth if a common channel was used. The channel assignment algorithm continues to assign the channels until all links in the network obtain appropriate channels. Upon the completion of this stage, the primary chromosome is obtained and the initial population can be created. The semi-chaotic technique is used to create the initial population from the primary chromosome. The weak genes (the link after the channel is assigned) with high interferences are replaced randomly from the available channels. Meanwhile, the strong genes (the links with non-overlapping channels) are fixed. In doing so, two main advantages are obtained that improve the metaheuristic approaches, namely the genetic algorithms for approaching the NP-Hard problems: the speed of the convergence and the effective solutions that are obtained. Because the semi-chaotic randomize a portion of gene space in the chromosome, the search scope is small, and thus fast convergence is obtained as well as the good quality of the obtained solution.

*C. Fairness-Aware Fitness Function*

After preparing the initial population using the semi-chaotic technique, the set of chromosomes in the initial population is evaluated. This required a fitness (objective) function in the genetic algorithm. The fitness function is used to evaluate the solution domain. As stated previously, the aim of the channel assignment is to maximize the link fairness while minimizing the link interference in order to maintain a suitable link data rate for the mesh clients so as to address node starvation problems. The objective function of the optimization is written as follows.

$$Maximize(Fairness\_index) \rightarrow (1)$$

Where $Fairness\_index$ can be expressed according to the Jain's index as follows.

$$Fairness\_index = \frac{[\sum_{i=1}^{n} link\_fairness_i]^2}{n \sum_{i=1}^{n} link\_fairness_i^2} \rightarrow (2)$$

Link fairness can be defined as a function of the link data rate as follows.

$$link\_fairness = \frac{Actual\ Link\ Data\ Rate}{Require\ Lin\ Data\ Rate} \quad \rightarrow (3)$$

The $link\_data\_rate$ needs to be maximized to achieve higher fairness. The *Actual Link Data Rate* is the expected data rate of the link after channel assignment while the *Require Lin Dat Rate* is the minimum accepted data rate of the link. In other words, the following condition should be satisfied.

$$Actual\ Link\ Data\ Rate \geq Require\ Lin\ Dat\ Rate \quad \rightarrow (4)$$

The $Required\_Link\_Data\_Rate$ can be calculated based on the sum of the required data rate of expected mesh clients or routers. The *Require Lin Dat Rate* of each link in the network is assumed to be known. This is a reasonable assumption as the number of clients that will be connected to the mesh routers can be controlled according to the available resources. The data rate is indirectly correlated with the interference i.e. if the interference is high the data rate is low and verse Versa. The actual data rate of a link can be calculated using Shannon–Hartley theorem which states that the data rate depends on the channel bandwidth and the signal to noise ratio, as in the following equation.

$$Actual\ Link\ Data\ Rate = BW * log_2(1 + SNR) \quad \rightarrow (5)$$

In the above equation, $BW$ is the bandwidth of the channel, $SNR$ is the signal-to-noise ratio, and capacity is the capacity of the channel in bits per second. To calculate the $SNR$, we first calculate the noise as the difference between the power of the transmitted signals and the power of the received signal.

Thus, the signal to noise ratio (SNR) can be calculated as follows.

$$SNR = \frac{(Power\ of\ Signal)}{(Power\ of\ Noise)} \quad \rightarrow (6)$$

SNR is usually expressed in decibels (dB) as follows.

$$SNR(db) = 10 * log_{10}\left(\frac{S}{N}\right) \quad \rightarrow (7)$$

As the bandwidth can be known from the channel characteristics and SNR can be calculated as the above equation, the data-rate can be calculated. Practically the strength of the received signal is measured using a hardware sensor, given that the transmission power is known; thus, the expected or the actual data rate after channel assignment can be estimated. However, for the sake of this study, the link data rate has been estimated using the following formula, which was derived from the Shannon-Hartley theorem and the path loss assuming Free-space path loss as follows.

$$RSS(dBm) = TSS(dBm) - 10 * n * log_{10}(link\_length)$$
$$RSS(dBm) - TSS(dBm) = -10 * n * log_{10}(link\_length)$$
$$TSS(dBm) - RSS(dBm) = 10 * n * log_{10}(link\_length)$$
$$Noise = 10 * n * log_{10}(link\_length)$$
$$\frac{TSS(dBm)}{Noise(dBm)} = \frac{TSS(dBm)}{10 * n * log_{10}(link\_length)}$$

$$SNR(dBm) = \frac{TSS(dBm)}{10 * n * log_{10}(link\_length)} \quad \rightarrow (8)$$

As the interference has an influence on the data rate, the interference on each length is the total overlapping of the link interference index.

$$SNR(dBm) = \frac{TSS(dBm)}{10 * n * link\_interferencence_l * log_{10}(link\_length)} \rightarrow (9)$$

Consequently, using this formula, the data rate can be estimated numerically. By minimizing the link interference index, the SNR will be higher, which increases the data rate as well. The link interference index can be estimated as follows.

$$link\ interferencence = \sum channel\ overlabing\ ratio_{l,m}\ \forall\ m\ \in l\ neighboring\ channels \quad \rightarrow (10)$$

The *channel overlabing ratio* can be estimated based on the conflicting graph as shown in Fig. 3 utilizing the interference matrix that was proposed in [9]. To granularly evaluate the fairness of each link in the network, the formula in Equation (3) was used. Then, by substitution, the $link\_fairness$ in the Jain's index equation which was presented in Equation (2), the following formula can be used to compute the overall fairness. Hence, the objective function can then be written as follows.

$$Fairness\ index = \frac{\left[\sum_{i=1}^{n} \frac{BW * log_2(1 + SNR)}{Require\ Lin\ Data\ Rate}_i\right]^2}{n \sum_{i=1}^{n} \left(\frac{BW * log_2(1 + SNR)}{Require\ Lin\ Data\ Rate}\right)_i^2} \rightarrow (11)$$

where the *BW* the channel bandwidth and *SNR* is the signal to noise ratio, which can be estimated using Equation 8. The aim of the genetic algorithm is to maximize the fairness index, as presented in Equation 11.

$$Max \begin{pmatrix} Fairness \\ index \end{pmatrix} \rightarrow (12)$$

### C. Fairness-Aware Parent Selection

Based on the fitness function presented in Equation 11, the individuals with the best fairness index are selected for the next population. The following formula is used to select the parents.

$$Selection\ Rule: if \begin{cases} Fairness\_index \geq |\mu + \sigma| & Selected \\ otherwise & Not\ Selected \end{cases} \rightarrow (13)$$

where $\mu$ is the mean of the fairness of the population, and $\sigma$ is the standard deviation of the population fairness. Although the number of selected parents is not necessarily fixed, the number of children (population size) is fixed in every generation.

### D. Semi-Chaotic Based New Generation

After selecting the parents, the mating process (crossover and mutation) is applied to produce the offspring of the next generation for the next iteration. The mating is the process of creating children from selected parents. This process consists of two main steps, namely crossover and mutation.

1. CROSSOVER: In the crossover process, the children are created from two parental chromosomes that were randomly selected from the previous step. The genes of the new child are mixed from the genes of the parents. The crossover process tends to pull the population towards the local maxima (the highest fairness in the population). For every two pairs, the powerful genes (high fairness genes) are fixed, and the weak genes (low fairness genes) are removed.
2. MUTATION: In the mutation step, the genes with low fairness are replaced randomly while the genes with high fairness are fixed. The mutation process aims at increasing the diversity and preventing the algorithm from being trapped in local minima.
   Fig 4 presents a flow chart of the proposed Fairness-Oriented Semi-Chaotic Genetic Based Channel Assignment (FA-SCGA-CAA)

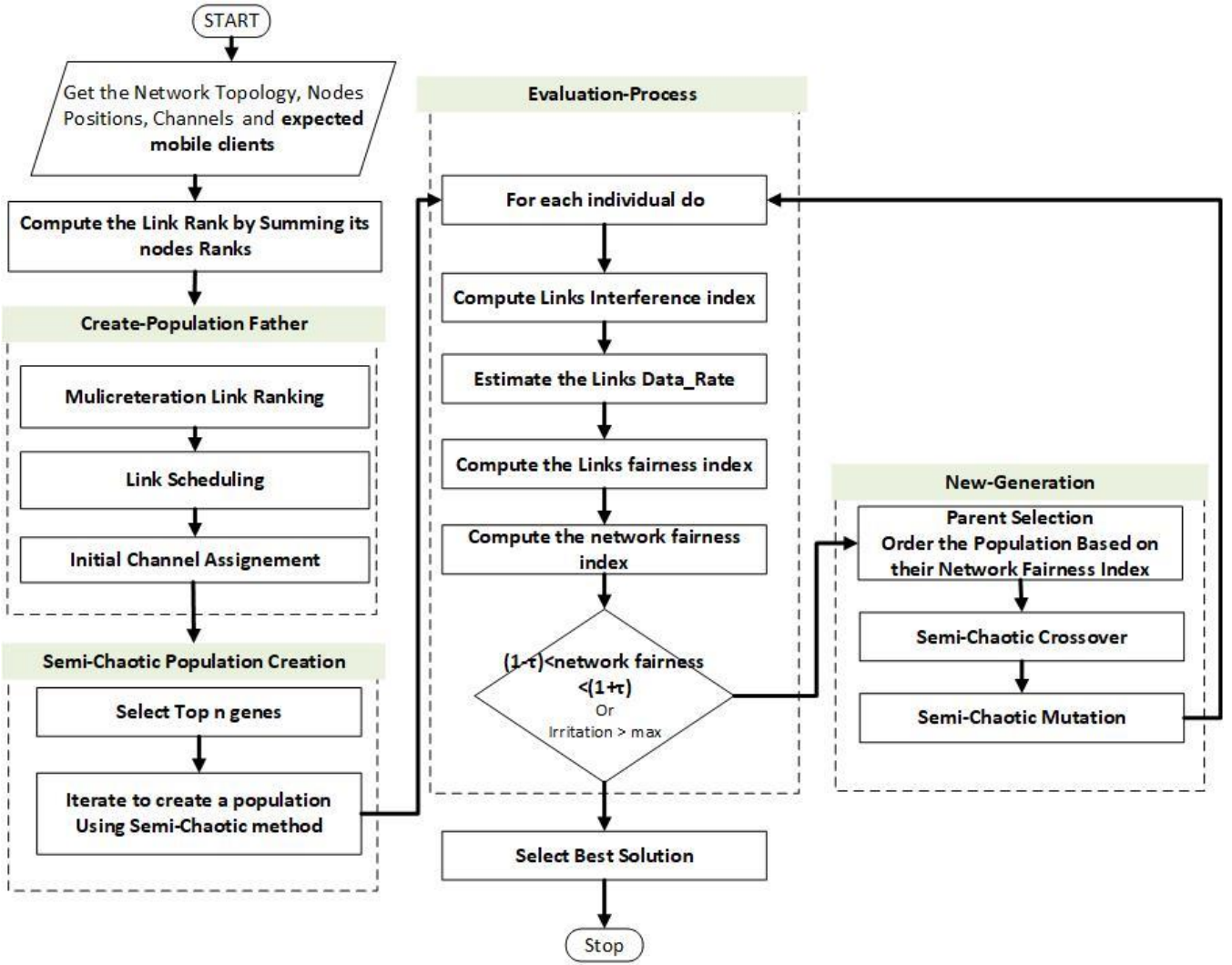

FIGURE 4. Flow chart of the proposed algorithm FA-SCGA-CAA

As shown in Fig. 4, the algorithm is divided into five phases as follows. The network representation phase was discussed in section A. In the network representation phase, the mesh network is represented by graph to facilitate the numerical analysis of the performance of the network. The primary chromosome formation and phase semi-chaotic based initial population creation were elaborated in section B. The fairness evaluation phase-was conducted using a novel fitness function that was developed to minimize the links interference while maximizing the fairness was presented in Section C. Natural selection was used to select the parents for the next generation. That is, the parents that have high fairness were selected. Meanwhile, parents with low fairness were neglected. The creation of the new generation phase was discussed in Section D. The subsequent sections evaluate and validate the proposed fair channel assignment solution.

## IV. PERFORMANCE EVALUATION

In this section, the performance of the proposed channel assignment algorithm is evaluated and compared with related work. The baseline algorithm starts assigning channels to nodes arranged in descending order based on the number of hops to the gateway [15, 19]. This algorithm may not provide a fair allocation of channels in the most sensible area of a WMN, like nodes far from the gateway. Since the direction of the traffic is towards the gateway, bottleneck problems may occur anywhere in that path, thereby causing network fragmentation, capacity degradation and the node starvation problem. Python's network library was utilized to implement the simulation of the algorithms presented in this paper.

To illustrate the performance of the proposed algorithm, five performance measures were used, namely, Network Capacity (NC), the Fractional Network Interference (FNI), per-link capacity, per-link fairness index, per-link interference index,. NC is the total concurrent transmission in the network after the channel assignment algorithm takes place, while FNI is the ratio between network interference and the total conflicting links in the network. FNI is also defined as the number of conflicts that remain after channel assignment relative to the number of conflicts in a single channel network. It is the remaining ratio of interference after applying the channel assignment algorithm. Jain's index [44] has been used to compute the fairness index. Jain's index is independent of the population size, scale and metric. It is bounded between zero and one where one indicates maximum fairness and zero indicates no fairness. Jain's index is calculated according as follows.

$$f(X) = \frac{[\sum_{i=1}^{n} x_i]^2}{n \sum_{i=1}^{n} x_i^2} \rightarrow (14)$$

where $n$ denotes the number of links, and $x_i$ is the link fairness assigned to the i[th] link as calculated in equation (2). It can be noticed from equation (14) that the fairness $f(X)$ can be 1 when $x_i$ value is equal for all. This indicates that every link has got fair channel allocation. The network throughput is the sum of the capacity of all links in the network [45]. Link capacity indicator can be calculated from the interference according to the following equation.

$$Link\_capacity_{l(i,j)} = \frac{1}{1 + interference_{l(i,j)}} \rightarrow (15)$$

$$Network\_capacity\ (NC) = \sum_{\forall\ l(i,j) \in L} capacity_{l(i,j)} \rightarrow (16)$$

where $l(i,j)$ is the link between nodes $i$ and $j$, $interference_{l(i,j)}$ is the interference on the link $l_{(i,j)}$ after channel, the assignment has taken place, and L is the number of available links in the network.

To analyze the impact of interference between the links, different scenarios were simulated, each of which has a different number of links, namely 5, 16, 36, 46, 58, 78, 119 and 126. With each scenario, many random topologies were generated and tested. All parameters used in this experiment were chosen according to the common practice in the field of channel assignment algorithms. Table 1 shows the parameters used in the simulation.

TABLE 1: Simulation Parameters

| Parameter | Configuration |
|---|---|
| Propagation Model | Free Space/Two Ray Ground |
| Antenna | Omni-direction |
| MAC Type | 802.11a/b |
| Orthogonal Channels | 12/3 |
| Communication Range | 252m |
| Interference Distance | 514m |
| Number of Nodes | Varies from 50 to 200 |
| Number of links | Varies from 10 to 130 |
| Number of Radios | 3 for each mesh router |
| Connectivity Degree | 3 |
| Simulation Area | 1000m x 1000m |

The interference model was used to estimate the interference between the links as follows. Two links interfere with each other if they have been set to the same channel and the distance between any two nodes that form the links smaller than the interference range (514m). The number of links was created autonomously to preserve network connectivity. That is, the links are created if the distance between the nodes smaller than the communication range (252m). For the generalization, in the simulation, the number of nodes, links, and their position have been randomly selected considering the connectivity in mind during nodes creations.

Two versions of the proposed algorithm were evaluated, namely the FA-SCGA-CAA and SCGA-CAA. FA-SCGA-CAA is the Semi-chaotic genetic algorithm with a fairness oriented fitness function which is the main contribution of this paper. Meanwhile, SCGA-CAA is the Semi-chaotic genetic algorithm without a fairness oriented fitness function. The fitness function used is the interference based fitness function. In other words, FA-SCGA-CAA aims at maximizing the fairness among nodes while minimizing the link interference while SCGA-CAA aims at minimizing the sum of link interference.

The proposed FA-SCGA-CAA and SCGA-CAA algorithms have been evaluated by comparing them with Multi Criterion Link Ranking CAA (MCLR-CAA) [9], and the Genetic Algorithm Based CAA (GA-CAA) that have been used for channel assignment in several related works [18, 19, 21].

## IV. RESULTS AND DISCUSSION

Extensive simulations were conducted to evaluate the performance of the proposed algorithms and techniques (FA-SCGA-CAA and SCGA_CAA). The results of the proposed SCGA_CAA algorithm were compared to the multi-criterion link ranking based channel assignment algorithm MCLR_CAA [9] and the interference-aware genetic algorithm-based channel assignment algorithm IA_GA _CAA that was frequently reported in the literature [18, 19, 21]. Fig. 5 illustrates the average capacity achieved by the studied channel assignment algorithms. The x-axis represents the average number of links in the studied network scenarios, while the y-axis represents the corresponding average network capacity. The network capacity is represented by the ratio of the concurrent connection in the network.

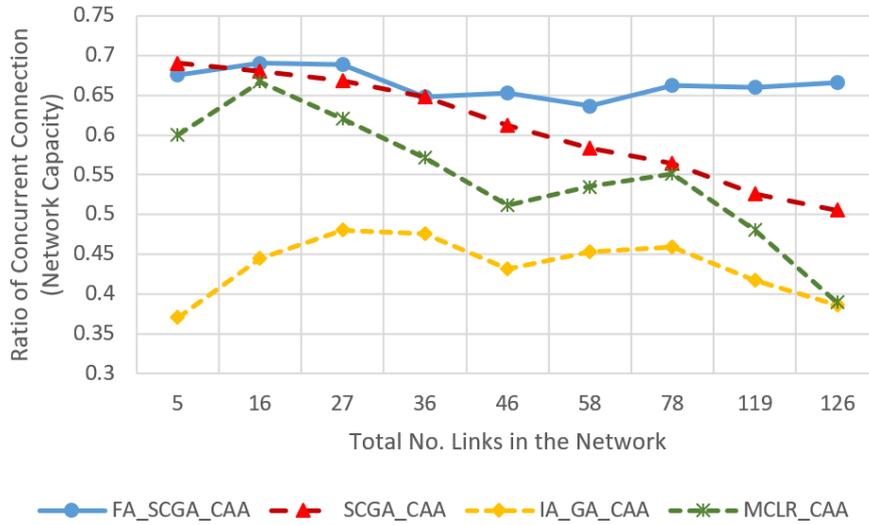

FIGURE 5. Network Capacity (NC)

As shown in Fig. 5, the fairness-oriented semi-chaotic algorithm FA_SCGA_CAA has the highest network capacity in all simulated scenarios compared with the other studied algorithms. It stays stable at more than 0.65 in most simulated scenarios. The proposed semi-chaotic algorithm SCGA-CAA archives higher network capacity than the related metaheuristic genetic algorithm based CAA (IA_GA_CAA) and the heuristic-based approach MCLR_CAA. MCLR_CAA, however, archives better network utilization than the metaheuristic algorithm (IA_GA_CAA). This is because the MCLR_CAA considers both interference and link fairness during channel assignment while the IA_GA_CAA tends to minimize the overall network interference without considering the fair distribution of the channels. This explains the improvement gained by the semi-chaotic approach SCGA-CAA when the MCLR_CAA was utilized to speed up the convergence towards better solutions. The proposed algorithm FA-SCGA-CAA outperforms the other studied algorithms in terms of network utilization. Considering node-fairness during channel assignment allows it to improve network performance. Unlike the multi-criterion algorithm (MCLR_CAA), the proposed FA-SCGA-CAA algorithm neither biases to specific links based on their criterion nor randomly picks a population that causes the algorithm trapping in local minima. FA-SCGA-CAA starts with a population containing a list of good solutions. Hence, it converges faster, and the results are more effective. It can be concluded that considering the link fairness during channel assignment not only prevents the node starvation problem but also improves the utilization of the available network capacity.

Figure 6 shows the achievements of the studied algorithms in terms of average Link Capacity. Per link capacity was calculated using equation (14) which expresses the capacity degradation due to the interference. The x-axis represents the average number of links in 9 scenarios, while the y-axis represents the corresponding average link capacity.

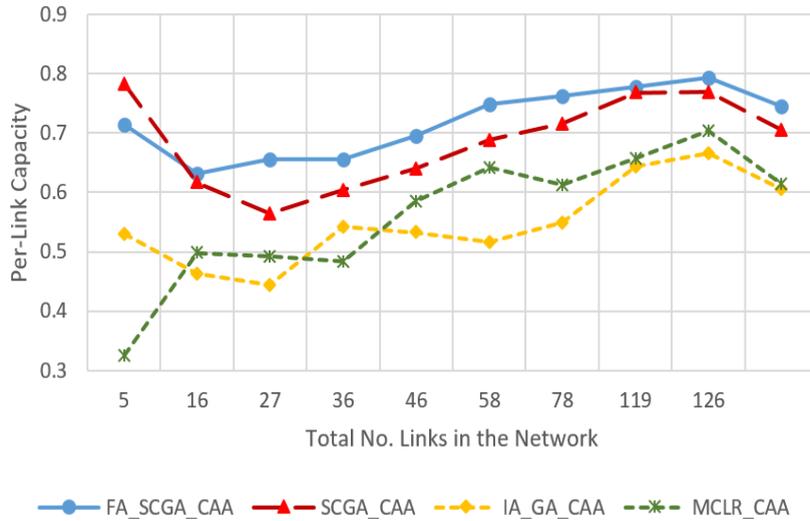

**FIGURE 6.** Per-Link Capacity

From Fig. 6, the proposed FA_SCGA_CAA achieves the highest average link capacity in most of the studied scenarios. The average link capacity slightly increases as the number of links increases in the network. It can be seen that the FA_SCGA_CAA outperforms the other studied algorithms under all studied scenarios. The proposed semi-chaotic based algorithm (SCGA_CAA) achieves better average link capacity than the multi-criterion link ranking algorithm (MCLR_CAA) and the interference-aware genetic algorithm (IA_GA_CAA). Although the FA_SCGA_CAA and SCGA_CAA show close achievements in terms of the average link capacity, they have different achievements in terms of network capacity (See Fig. 5). The network capacity of SCGA_CAA drops rapidly as the number of links increases while the network capacity of FA_SCGA_CAA slightly decreases. It can be noticed that in all studied algorithms, the average link capacity increases with increasing the number of links in the network. This can be interpreted as follows. As the network grows bigger, the coverage distance becomes wider which makes the interference become smaller. It increases the number of links with non-overlapping channels, and so does the average link capacity.

Fig. 7 illustrates the achievements of the studied algorithms in terms of the average interference index per link. The link interference index is represented by the number of the overlapping neighboring links. The x-axis in Fig. 7 represents the average number of links of the 9 studied scenarios, while the y-axis represents the corresponding average link interference.

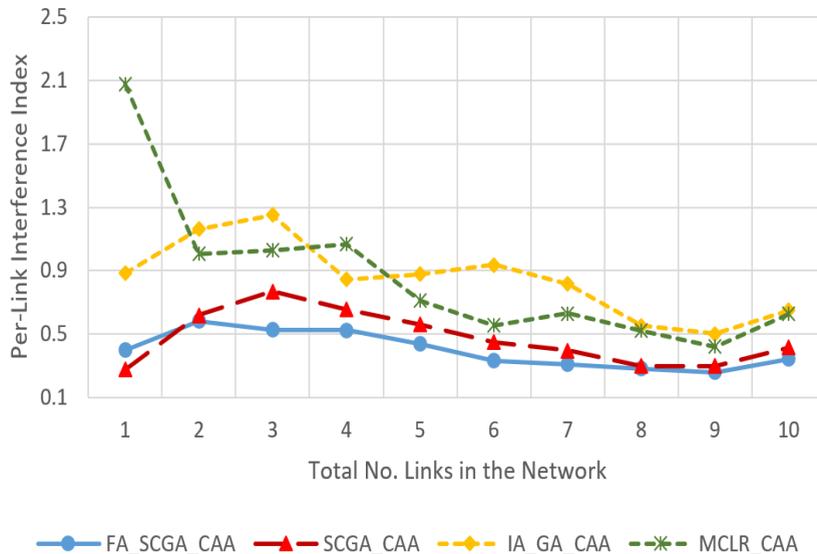

**FIGURE 7.** Per-Link Interference

As shown in Fig. 7, the proposed FA_SCGA_CAA achieved the lowest average link interference among the studied algorithms. It can be noted that in all the studied scenarios, the average link interference decreases as the size of the network increases due to the sparse nature of large networks where the number of links with non-overlapping channels increases.

Figure 8 presents the fractional network interference (FNI). FNI measures the potential amount of nodes that exposed to the starvation problem, and calculated based on the number of conflicted channels in the network. In Figure 8, the x-axis represents the average number of links in the 9 studied scenarios, while the y-axis represents the fractional network interference (FNI).

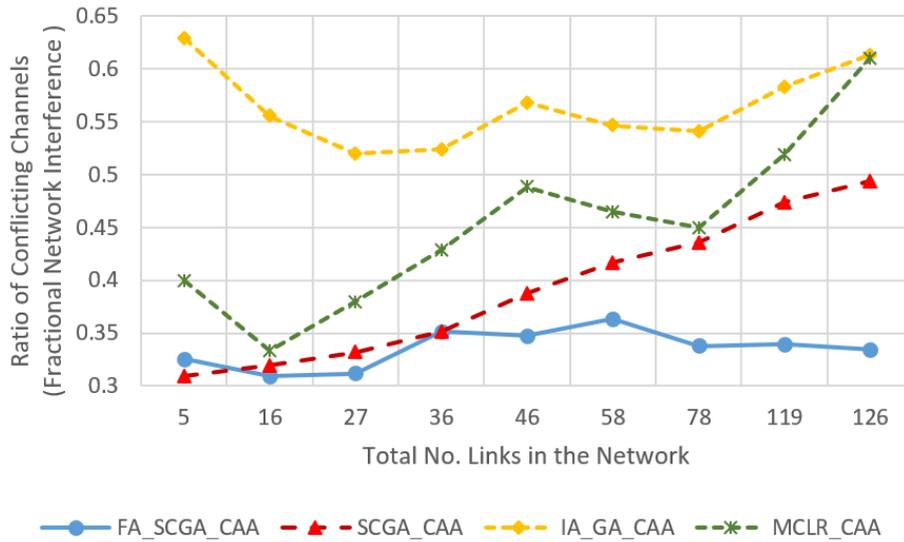

**FIGURE 8.** The fractional network interference

From Fig. 8, it can be seen that the FNI of the proposed FA_ SCGA-CAA remains stable below than 35% with all studied scenarios. It slightly fluctuates between 30% and 35%, which implies that the ratio of the potential node starvation will be as lower as 35% in the worst cases. The FNI of the semi-chaotic genetic algorithm (SCGA-CAA) and the multi-criterion link ranking algorithm (MCLR-CAA) increases as the amount of links increases. Meanwhile, the FNI of the interference-aware genetic algorithm (IA_GA-CAA) remains stable at above 50% in all studied scenarios. This is because IA_GA-CAA does not consider the fair distribution of the channels among available links while it is partially considered in both SCGA-CAA and MCLR-CAA. It can be concluded that the multi-criterion-based algorithm (MCLR-CAA) and the semi-chaotic genetic-based algorithm (SCGA-CAA) can reduce the link interference and thus, nodes starvation.

Fig. 9 illustrates the achievements in terms of average link fairness. The link fairness is calculated according to Equation (10). The fairness is the ratio between the link data rate after channel assignment and the link capacity before the channel assignment. The maximum fairness index is between one (for the fair distribution) and zero (for the unfair distribution). The x-axis in Fig. 9 represents the average number of links in the 9 studied scenarios while the y-axis represents average link fairness.

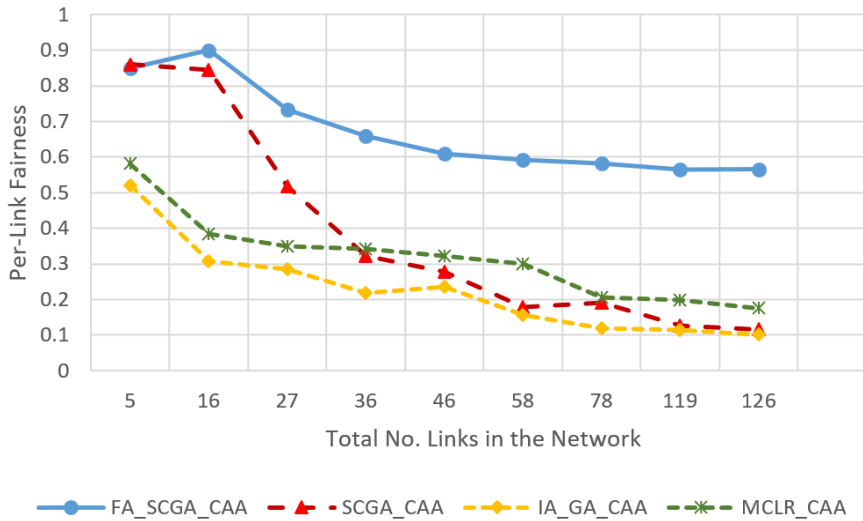

**FIGURE 9.** Per-Link Fairness

From Fig. 9, the proposed FA_SCGA_CAA achieves the highest average fairness index among all studied algorithms and in all the simulated scenarios. The fairness index slightly drops as the number of links increases. Although the fairness index of the semi-chaotic genetic algorithm (SCGA-CAA) is high when the number of links is low, it rapidly drops from 0.85 until 0.1 when the number of links increases. Meanwhile, the fairness indexes of the MCLR-CAA and IA_GA-CAA are low and drop lower than 0.2 when the network grows bigger.

From Fig. 9, 8, 7, 6, and 5, it can be concluded that the developed fairness fitness functions that maximize the fairness and minimize the interference not only prevented node starvation problem, but also improves the utilization of network capacity. Fig. 10 presents a comparative summary of the performance of the proposed FA_SCGA_CAA and the SCGA-CAA, and the related works MCLR-CAA, and IA_GA-CAA. The x-axis represents the performance measures, namely the average of the average network capacity in terms of concurrent connection, the average of potential starvation nodes in terms of the ratio of the conflicting channels or the fractional network interference (FNI), the average link fairness index, and network interference index. The y-axis represents the average value of the corresponding performance measures.

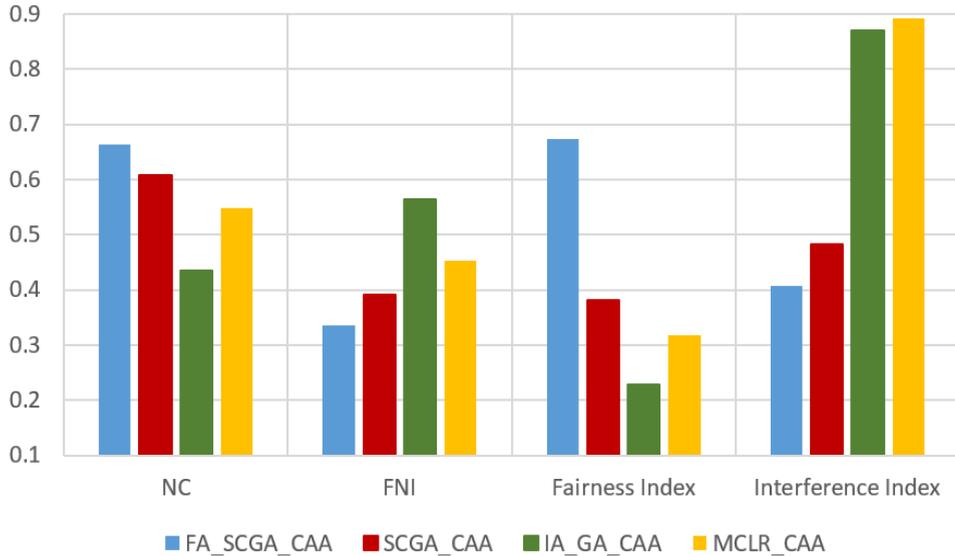

**FIGURE 10.** Summary of the overall performance of the proposed SCGA_CAA and related work.

As shown in Figure 10, the proposed FA_SCGA_CAA outperforms the other studied algorithms with respect to every analyzed performance measure. In terms of network capacity, the FA_SCGA_CAA algorithm improves the utilization of the network capacity by 23% comparing by IA_GA_CAA. Likewise, FA_SCGA_CAA also reduces the potential nodes starvation by 22% as compared with IA_GA_CAA. In terms of average link fairness, the proposed FA_SCGA_CAA algorithm improves the fairness by 44% comparing by the fairness unaware approach IA_GA_CAA. In terms of interference 46% reduction is achieved by the FA_SCGA_CAA as compared with the IA_GA_CAA.

Fig. 11 illustrates the performance gained by applying the semi-chaotic approach in speeding up the convergence rate of the genetic algorithm. The x-axis represents the average number of links in the 9 studied scenarios while the y-axis represents the average iterations until convergence.

From Fig 11, the proposed semi-chaotic method has improved the speed rate of the convergence as compared with the original genetic algorithms. This is because the semi-chaotic method characterized the selected search space by more powerful genes. Hence, selecting powerful primary chromosome leads to create a population with powerful genes that speeds up the convergence. Furthermore, the multi-criterion method that is used to create the primary chromosome considers both the fairness and the interference during link ranking. Thus, it directs the convergence towards the global optima. This explains the improvements gained by the proposed FA_SCGA_CAA which employs both the fairness-based fitness function and the semi-chaotic approach to create the primary chromosome and search for the most effective solution.

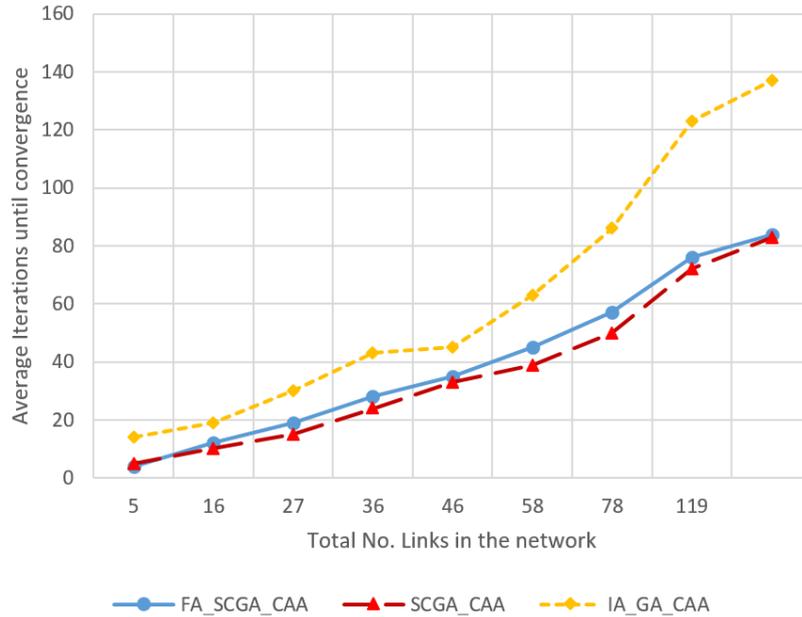

FIGURE 11: Summary of the overall performance of the proposed SCGA_CAA and related work.

## VI. CONCLUSION AND FUTURE WORK

In this paper, the Fairness-Oriented Semi-Chaotic Genetic Algorithm-Based Channel Assignment Technique (FA-SCGA-CAA) for Client Starvation Problem in Wireless Mesh Network was proposed. The nodes starvation problem that was overlooked by the extant studies due to unfair channel assignment has been addressed in this study. In order to achieve client fairness, a semi-chaotic genetic algorithm-based technique was proposed to create diverse population with informative features that converges to the best solution and avoids being trapped in the local minima. The channel assignment problem was formulated as an optimization problem with a fairness aware fitness function. The fairness oriented fitness function combines several factors that represent the network topology, load status and required bandwidth/throughput in one function. The fairness was defined on node level to address the node starvation problem that is overlooked by existent research. Extensive experimental evaluations were conducted to measure the performance of the proposed technique and compare it with existing solutions. The results showed that the proposed algorithm outperformed existing solutions in terms of improving the link fairness and utilization of network capacity while reducing the interference and the potential number of nodes starvations.

Similar to any genetic-based algorithm, the proposed algorithm has two main drawbacks as follows. The first issue is that setting up the number of iteration is scenario-specific. Therefore, it is difficult to select a fixed number of iteration for generalization. The second issue is on the selection of the convergence threshold. These issues contributed to the overall link fairness. We are currently working on addressing those two issues and the new findings will be the subject of our next publication.

## ACKNOWLEDGMENT
This work was supported by the Ministry of Higher Education (MOHE) and Research Management Centre (RMC) at the Universiti Teknologi Malaysia (UTM) under Post-Doctoral Fellowship Scheme (VOT R.J130000.7828.4F809).

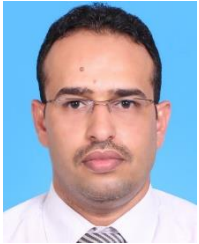
Fuad A. Ghaleb received his B.Sc. degree in Computer Engineering from Sana'a University, Yemen, 2003. He received the M.Sc. and Ph.D. degrees in computer science - information security from Faculty of Computing, Universiti Teknologi Malaysia, Johor, Malaysia, 2014, and 2018 respectively. From 2004-2012, he was lecturer of network and computer engineering at Sana'a Community College, Yemen. He is the author of 22 articles related to information and network security. His research interest includes Vehicular Network Security, Cyber threat intelligence, Intrusion Detection, Data Science, Data Mining, and Knowledge Discovery. Dr. Fuad was a recipient of the many awards and recognitions such as Post-Doctoral Fellowship award, Best Postgraduate Student Award, and Best Presenter Award from School of Computing, Faculty of Engineering, UTM, Malaysia as well as Best Papers Awards from IICIST, Kuala Lumpur, Malaysia and Effat University, Jeddah, Saudi Arabia.

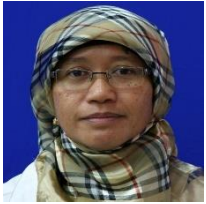
Maznah Kamat received her B.Sc degree in Computer Science from Western Michigan University, Kalamazoo, Michigan, U.S.A., M.Sc degree in Computer Science from Central Michigan University, Mount Pleasant, Michigan, U.S.A., and the Ph.D. degree in IT Security from Aston University, Birmingham, UK.
Dr. Maznah Kamat is a senior lecturer at the Faculty of Computing and the research dean of Smart Digital Community Research Alliance, Universiti Teknologi Malaysia (UTM). His research interest is in Information System Security. He is also the head of UTM-CSM Cyber Security X Lab and a member of Information Assurance & Security Research Group (IASRG), UTM.

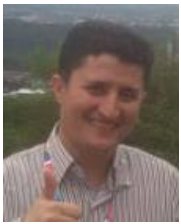
Bander Ali Saleh Al-rimy received his B.Sc degree in Computer Engineering from Faculty of Engineering, Sana'a University, Yemen, and M.Sc (Information Technology) from OUM, Malaysia, and the Ph.D. degree in Computer Science - Information Security from Faculty of Engineering, Universiti Teknologi Malaysia (UTM). Dr. Bander is a recipient of several academic awards, such as Thesis Merit Award, Excellence Award, Best Research Paper Award, and Distinction Award. His research interest includes but not limited to Malware, IDS, Network Security, and Routing Technologies.

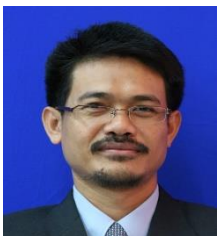
Mohd Fo'ad Rohani received his B.E (Hons.) and M.Sc. degrees in Electrical and Electronic Engineering from University Malaya (UM), Kuala Lumpur, and University of Wales, Cardiff UK in 1994 and 1998, respectively. In 2013, he obtained his Ph.D. (Computer Science) degree from Universiti Teknologi Malaysia (UTM) in the area of Network Security and Pattern Recognition. Currently he is a Lecturer in the Department of Computer Science, Faculty of Engineering, School of Computing, UTM. He is member of IEEE, Malaysia Section, and also a member of Information Assurance and Information Security Research Group (IASRG). His research interests include Pattern Recognition, Digital Signal Processing (DSP), Computer Architecture, Network Communication, and Security.

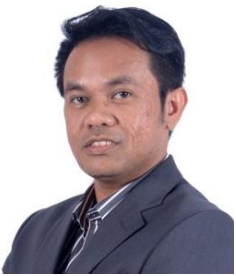
**Shukor Abdul Razak** is currently an Associate Professor at Universiti Teknologi Malaysia. He has authored and co-authored many journals and conference proceedings at national and international levels. His research interests are on the security issues for mobile ad hoc networks, mobile IPv6, vehicular ad hoc networks, and network security. Assoc.Prof. Dr.Shukor actively conducts several types of research in digital forensic investigation, wireless sensor networks, and cloud computing.